\documentclass[prl,twocolumn,showpacs,lengthcheck]{revtex4-1}

\pdfoutput=1

\usepackage{amsmath}
\usepackage{amssymb}
\usepackage{graphicx}

\newcommand{\bra}[1]{\left\langle{#1}\right\vert}
\newcommand{\ket}[1]{\left\vert{#1}\right\rangle}

\newcommand{\tr}{\operatorname{tr}}

\begin{document}

\bibliographystyle{apsrev}

\title{Bounds on Information Propagation in Disordered Quantum Spin Chains}
\author{Christian K. \surname{Burrell}}
\affiliation{Department of Mathematics, Royal Holloway University of
London, Egham, Surrey, TW20 0EX, UK}

\author{Tobias J. \surname{Osborne}}
\affiliation{Department of Mathematics, Royal Holloway University of
London, Egham, Surrey, TW20 0EX, UK}

\begin{abstract}
We investigate the propagation of information through the disordered
$XY$ model. We find, with a probability that increases with the size of the system, that all correlations, both classical and
quantum, are suppressed outside of an effective
lightcone whose radius grows at most polylogarithmically with $|t|$.
\end{abstract}


\maketitle

How fast can information propagate through a locally interacting
system? For classical systems an essentially universal answer to
this question is that the velocity of information propagation is
bounded (often only approximately) by an effective \emph{speed of
light}. It is a more subtle issue to formulate equivalent velocity
bounds for quantum systems because they can encode \emph{quantum
information} in the form of \emph{qubits} and therefore might be
able to exploit quantum interference to propagate information
faster. However, for quantum spin networks this is not the case: the
\emph{Lieb-Robinson bound} limits the velocity at which correlations
can propagate \cite{lieb:1972a}.

The Lieb-Robinson bound implies that there is an effective light
cone for two-point dynamical correlations, i.e., apart from an
exponentially suppressed tail, two-point correlations propagate no
faster than a speed of light. Simplified and alternative proofs of
the Lieb-Robinson bound have been subsequently discovered
\cite{nachtergaele:2005a, hastings:2004a, hastings:2005b,
osborne:2006b}. More recently, it has been realised that the
Lieb-Robinson bound is strong enough to bound not only the
propagation of two-point correlations but of any local encoding of
information: regardless of the encoding no information (either
quantum or classical) can propagate faster than the speed of light
for the system \cite{bravyi:2006a}.

There are many consequences of the Lieb-Robinson bound. Apart from
the aforementioned bounds on the velocity of information
propagation, it has been realised that the Lieb-Robinson bound can
be used to provide a method to efficiently simulate the properties
of low-dimensional spin networks \cite{osborne:2005d, osborne:2005c,
osborne:2006a, osborne:2006c, hastings:2006a}. Additionally, using
the Lieb-Robinson bound, dynamical entropy area laws for quantum
spin systems can be obtained \cite{bravyi:2006a, eisert:2006b}.

\begin{figure}
\center
\includegraphics{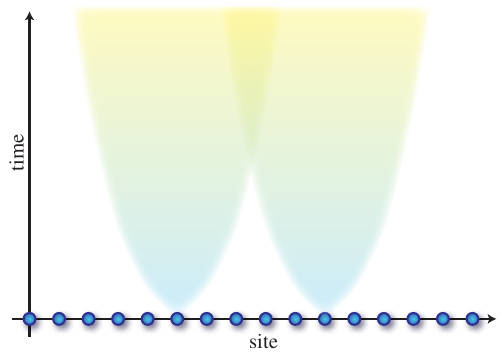}
\caption{Schematic illustration of the conjectured \emph{logarithmic
light cone} for disordered systems: as time progresses information
is exponentially attenuated outside of a light cone whose radius
grows at most logarithmically with time.}\label{fig:loglightcone}
\end{figure}

While the Lieb-Robinson bound is extremely general --- it relies
only on the ultraviolet cutoff imposed by lattice structure --- it
is, as a consequence, relatively weak. Thus, it is extremely
desirable to develop stronger bounds constraining the propagation of
quantum information through systems where more is known about the
structure of the interactions. One setting where one would expect
stronger bounds to be available is that of a quantum spin system
with disordered interactions. Such systems have attracted a large
amount of interest as they can exhibit the striking phenomenon of
\emph{Anderson localisation} \cite{anderson:1958b}, which means that
information is essentially frozen: a quantum particle placed
anywhere within a localised system diffuses only slightly, even for
extremely large times. Thus, exploiting the parallels between bounds
on information propagation and Lieb-Robinson bounds, we are
motivated to conjecture that interacting spin systems with
disordered interactions satisfy stronger bounds on correlation
propagation (see Fig.~\ref{fig:loglightcone}). More specifically, we
conjecture that for quantum spin networks with disordered
interactions all correlations, both quantum and classical, are
suppressed outside of a light cone whose radius grows at most
\emph{polylogarithmically} in time. (Contrast this with the light cone
supplied by the Lieb-Robinson bound: it has a radius which grows
linearly with time.)

In this Letter we study a setting where the dynamics of a class of
disordered interacting spin systems can be shown to satisfy our
polylogarithmic light-cone conjecture; we study the $XY$ spin chain with
disordered interactions in a disordered magnetic field and show that
information, and hence correlations, are attenuated
outside of a light cone whose radius grows polylogarithmically with
time. The main result of this Letter, a \emph{polylogarithmic light
cone} for the disordered $XY$ model, can by summarised with the
following bound on the dynamic two-point correlation functions:
  \begin{equation} \label{ijbound}
    \left\| \left[ A_j , e^{itH} B_k e^{-itH} \right]
    \right\| \leq c_\zeta n^2  |t| e^{-\gamma_\zeta|j-k|^\zeta},
  \end{equation}
which holds for any $0<\zeta<1$ with $n$ larger than a constant depending on $\zeta$ and $\gamma_\zeta$, with probability $p \ge 1-e^{-n^\zeta}$,
where $A_j$ and $B_k$ are local operators acting nontrivially only
on spins $j$ and $k$ respectively, and $\gamma_\zeta$ and $c_\zeta$ are constants proportional to $\zeta$ and the second moment of the disorder distribution. We apply our new bound to study the structure of the
propagator for large times and the scaling of the entropy of a block
of spins in the evolving system. As a consequence, we prove the
entropy saturation numerically observed by De Chiara \emph{et.\ al.}
\cite{dechiara:2006a}. Our results also constitute a proof of a
conjecture raised in \cite{keating:2006a}: namely, if two parties,
Alice and Bob, have access to a bounded region at either end of the
chain, respectively, then it is impossible for Alice to send any
information to Bob, regardless of how Alice encodes the information
in the single- and higher-excitation sectors.

We consider a one-dimensional chain of $n$ spin-$1/2$ particles with
$XY$-model type interactions between nearest-neighbouring spins in
an additional transverse field (e.g.\ a magnetic $B$-field). We
allow the coefficients of the couplings and the transverse field
strength to vary from site to site within the spin chain. Thus, we
study the evolution of the chain under the Hamiltonian
  \begin{equation} \label{hamiltonian}
    H = \sum_{j=1}^{n-1} \mu_j \left( \sigma_j^X\sigma_{j+1}^X +
    \sigma_j^Y\sigma_{j+1}^Y \right) + \sum_{j=1}^{n}
    \nu_j\sigma_j^Z,
  \end{equation}
where $\mu_j$ and $\nu_j$ are drawn from probability distributions
$\mathbb{P}_\mu$ and $\mathbb{P}_\nu$ respectively, and where
$\sigma_j^\alpha$ ($\alpha \in \{X,Y,Z\}$) is a Pauli spin operator
acting on the spin at site $j$. Typically, $\mu_j = -J$ for all $j$,
however this is not necessary and we deal with the more general case
here.

We solve this system using the \emph{Jordan-Wigner transform} (for
an introduction to the Jordan-Wigner transform see
\cite{nielsen:2005a}) which, when combined with some exact results
from the theory of localisation, allows us to bound the dynamics of
our spin chain. We note that it is not immediate that the dynamics
of the disordered $XY$ model satisfy a logarithmic light cone: while
the $XY$ model is equivalent to a free fermion model which is the
original Anderson model, and is localised, the Jordan-Wigner
transform is a \emph{nonlocal} operation and might confound the
localisation occurring in the free fermion system.

Let's start by applying the Jordan-Wigner transform, which maps a
system of interacting qubits into a system of free fermions. The
Jordan-Wigner transform defines annihilation operators $a_j =
(\sigma^Z_1 \cdots \sigma^Z_{j-1}) \sigma_j$ (where $\sigma_j =
\ket{0}\bra{1}$ acts on site $j$) and the corresponding creation
operators $a^\dag_j$, which satisfy the canonical fermionic
anticommutation relations. Using this we are able to rewrite the
system Hamiltonian as $H = \sum_{j,k=1}^n M_{jk} a_j^\dag a_k$,
where the tridiagonal matrix $M$ is defined via $M_{j,k} = 2\mu_k
\delta_{j,k+1} + 2\mu_j \delta_{j,k-1} - 2\nu_j \delta_{j,k}$.

It is now possible (following the method described in
\cite{nielsen:2005a}) to diagonalise $H$. After doing so we find the
dynamics for the annihilation operators in the Heisenberg picture,
with $a_j(t) = e^{itH} a_j e^{-itH}$:
  \begin{equation}\label{ajt}
    a_j(t) = \sum_{k=1}^n v_{jk}(t) a_k,
  \end{equation}
where $v_{jk}(t) = \left( e^{-iM t} \right)_{j,k}$. We now
concentrate on bounding the $v_{jk}(t)$, which in turn bounds the
dynamics of the system.

The quantity $v_{jk}$ has been well studied in the physical and mathematical literature. At the level of physical rigour it is typically argued that $v_{jk}$ decays, with probability increasing with $n$, exponentially with separation. That is, $|v_{jk}| \le c e^{-\gamma|j-k|}$, where $c$ and $v$ are constants depending only on $\gamma$, a constant proportional to the (assumed finite) \emph{second moment} of the probability distribution $\mathbb{P}_\nu$ (which we assume is i.i.d.). The mathematical literature hasn't yet achieved results as good as this (although this situation is recently changing, see \cite{chulaevsky:2011a} for recent progress). Instead, the best currently available result is obtained via \emph{bootstrap multiscale analysis} (see, e.g., \cite{germinet:2001a}), and reads
\begin{equation}\label{vijbound}
	|v_{jk}| \le c_\zeta e^{-\gamma_\zeta|j-k|^\zeta},
\end{equation}
which holds for any $0<\zeta<1$ when $n$ is greater than a constant depending only on $\gamma_\zeta$, with probability $p \ge 1-e^{-n^\zeta}$, where $c_\zeta$ and $\gamma_\zeta$ are constants depending only on $\zeta$ and the second moment of the disorder distribution. These results are typically obtained for infinite lattices, however the proof technique may be adapted to show the result in the finite-size case that concerns us here \cite{klein:2011a}.

The inequality
Eq.~(\ref{vijbound}) is a quantitative statement of the result that
the modulus of the diagonal matrix elements of $e^{-iM t}$ are
large, while the modulus of the off diagonal matrix elements decay
 with distance from the diagonal. This means that
$a_j(t)$ is effectively a linear combination of only a small number
of $a_k$ operators --- namely those for which $|j-k|$ is small. It
is this fact which inhibits the spread of operators on the chain,
giving rise to the light cone we derive below.

We now turn to the proof of the improved Lieb-Robinson bound for our
system. We begin by bipartitioning the spin chain into two sections,
$A$ and $B$, where we assume the boundary between partitions is
between spins $m$ and $m+1$. We then attempt to write $e^{itH}$ as a
product of $e^{itH_A}$ and $e^{itH_B}$. Clearly this won't be exact
and so we introduce an operator $V(t)$ which bridges the boundary
between $A$ and $B$, and which is designed to compensate for any
errors introduced:
\begin{equation}
  e^{itH} = e^{it(H_A + H_B)}V(t).
\end{equation}
The operator $V(t)$ acts nontrivially on all spins in the chain,
however, we now show that $V(t)$ can be well approximated by another
operator, which we call $V^\Omega(t)$, which acts only on a small
number $|\Omega|$ of spins. The reason we can do this is that $V(t)$
acts strongly on spins which are close to the boundary and
progressively weaker on spins as we move away from the boundary. To
prove this approximation is valid, we use the following differential
equation for $V(t)$:
\begin{equation}
\frac{dV}{dt}  =  i V(t)h_m(t),
\end{equation}
where $h_m(t) = e^{-itH} h_m e^{itH}$ and $h_m$ is the interaction
term in the Hamiltonian which bridges the boundary. We let $\Omega$
denote a set of $|\Omega|$ spins centred on the boundary between the
partitions $A$ and $B$. We also define $h^\Omega_m(t) =
e^{-itH_\Omega} h_m e^{itH_\Omega}$ where $H_\Omega$ contains only
those interactions in $H$ which act on sites in $\Omega$. We then
define $V^\Omega(t)$ via
\begin{equation}
  \frac{d}{dt}V^\Omega(t) = i V^\Omega(t) h^\Omega_m(t).
\end{equation}
Clearly the operator $V^\Omega(t)$ acts nontrivially only on
$\Omega$.

The error between $V(t)$ and $V^\Omega(t)$ is bounded by
\begin{equation}
  \left\|V(t) - V^\Omega(t)\right\| \leq \int_{0}^{|t|} \left\|
  h_m(s) - h^\Omega_m(s) \right\| ds.
\end{equation}
Calculating $\|h_m(t) - h^\Omega_m(t)\|$ is a lengthy but
straightforward task, and we begin by using the Jordan Wigner
transform to write this quantity in terms of the $a_j$ operators:
\begin{multline}
  h_m  =  2\mu_m \left( a^\dag_m a_{m+1} - a_m a^\dag_{m+1}
  \right) + \frac{\nu_m}{2} \left(a_m a^\dag_m - a^\dag_m a_m
  \right) \\
  + \frac{\nu_{m+1}}{2} \left( a_{m+1} a^\dag_{m+1} -
  a^\dag_{m+1} a_{m+1} \right).
\end{multline}
When we calculate $\|h_m(t) - h^\Omega_m(t)\|$ we'll have to deal
with terms such as $\|a^\dag_m(t) a_{m+1}(t) - a^{\Omega \dag}_m(t)
a^\Omega_{m+1}(t)\|$, which can be bounded as follows
  \begin{multline}
  \left\| a^\dag_m(t) a_{m+1}(t) - a^{\Omega \dag}_m(t)
    a^\Omega_{m+1}(t) \right\|
    \leq \\ \left\| a^\dag_m(t) \right\| \left\| a_{m+1}(t) -
    a^\Omega_{m+1}(t) \right\| \\ + \left\| a^\dag_m(t) - a^{\Omega \dag}_m(t) \right\|
    \left\| a^\Omega_{m+1}(t) \right\|.
  \end{multline}
Now $\|a_m(t)\| = \| a^\Omega_m(t)\| = 1$, so we've reduced the
problem of bounding $\|h_m(t) - h^\Omega_m(t)\|$ to bounding
$\|a_m(t) - a^\Omega_m(t)\|$. The operator $a^\Omega_m(t)$ is given
by
\begin{equation}
  a^\Omega_m(t) = \sum_{k\in\Omega} v_{mk}(t) a_k.
\end{equation}
Hence we arrive at
\begin{equation}
  \left\| a_m(t) - a^\Omega_m(t) \right\| = \left\|
  \sum_{k\notin\Omega} v_{mk} a_k \right\|\le \sum_{k\notin\Omega}
  nc_\zeta e^{-\gamma_\zeta|m-k|^\zeta},
\end{equation}
where we've used our bound Eq.~(\ref{vijbound}) on $|v_{jk}|$ and
the fact that $\|a_k\| = 1$.

Since $\Omega$ is a set centred on the boundary between partitions
$A$ and $B$ of the chain, we have that $|m-k| \geq |\Omega|/2$ for
all $k\notin\Omega$. Hence
\begin{equation}
  \left\| a_m(t) - a^\Omega_m(t) \right\| \leq
  c_\zeta n\left(n-|\Omega|\right) e^{-\gamma_\zeta|\Omega|^\zeta/2}
\end{equation}
and so we are finally able to conclude that $\left\| h_m(t) -
h^\Omega_m(t) \right\| \leq c n^2 e^{-\gamma_\zeta|\Omega|^\zeta/2}$ and that
\begin{equation}
  \left\| V(t) - V^\Omega(t) \right\| \leq c_\zeta|t|n^2
  e^{-\gamma_\zeta|\Omega|^\zeta/2},
\end{equation}
(here we have redefined the constant $\gamma_\zeta$). In particular, given $\epsilon \geq 0$,
choosing $|\Omega|^\zeta \geq 2\log{(c_\zeta|t|n^2/\epsilon)}/\gamma_\zeta$ ensures
that $\left\| V(t) - V^\Omega(t) \right\| \leq \epsilon$. Even $\zeta=1/2$ gives a polylogarithmic light cone whose width grows as the square of a logarithm of $|t|$. This may be improved arbitrarily by choosing larger $\zeta$ at the expense of worse constants. That is,
given any $\epsilon \geq 0$ we can choose $\Omega$ to be a large
enough set such that $V^\Omega(t)$ approximates $V(t)$ to within
$\epsilon$. This enables us to write
\begin{equation}
  e^{itH} = e^{it(H_A + H_B)} V^\Omega(t) + \mathcal{O}(\epsilon)
\end{equation}
Following \cite{osborne:2005d} we recursively apply the above
partitioning procedure to find $e^{itH} = Q(t) +
\mathcal{O}(\epsilon)$, where
\begin{equation}\label{e_ith}
  Q(t) \equiv \left( \bigotimes_{j=1}^{n/|\Omega|}
  e^{itH_{\Omega_j}} \right) \left(
  \bigotimes_{k=0}^{n/|\Omega|} V^{\Omega^\prime_k}(t) \right),
\end{equation}
and where $\mathcal{P}_1 = \{\Omega_j\}$ is a partition of the chain
into $\frac{n}{|\Omega|}$ blocks, each containing $|\Omega|$ spins
and where $\mathcal{P}_2 = \{\Omega^\prime_k\}$ is a partition of
the chain obtained by shifting $\mathcal{P}_1$ along by
$\frac{|\Omega|}{2}$ sites (note that $\Omega^\prime_0$ and
$\Omega^\prime_{n/|\Omega|}$ are half-size blocks of
$\frac{|\Omega|}{2}$ sites each). This is our fundamental structure
result for the dynamics of the disordered $XY$ spin chain.

A Lieb-Robinson bound is an upper bound on quantities such as $\| [
A, B(t)]\|$. We now show how the above structure result implies a
version of the Lieb-Robinson bound which is substantially stronger
than the original. Define $\widetilde{B}(t)$ to be the operator
which arises when we evolve $B$ according to the approximation
$Q(t)$ of $e^{itH}$, namely, $\widetilde{B}(t) = Q(t)BQ^\dag(t)$.
This enables us to write $B(t) = \widetilde{B}(t) +
\mathcal{O}(\epsilon)$. Note that $\widetilde{B}(t)$ acts trivially
on all sites which are a distance greater than $3|\Omega|/2$ away
from those sites on which $B$ acts. Thus, if $d(A,B)\geq
3|\Omega|/2$, where $d(A,B)$ is the distance between $A$ and $B$,
then $[A,\widetilde{B}(t)]=0$, and so for a given $|\Omega|$:
\begin{eqnarray}
  \left\| \left[ A, B(t) \right] \right\| & = &
  \left\| \left[ A, \widetilde{B}(t) \right] +
  \left[ A, \mathcal{O}(\epsilon) \right] \right\| \nonumber \\
  & \leq & 2 \left\| A \right\| \left\| \mathcal{O}(\epsilon) \right\|
  \nonumber \\
  & \leq & c_\zeta n^2 |t| e^{-\gamma_\zeta|\Omega|^\zeta/2}\\
  & \leq & c_\zeta n^2 |t| e^{-\gamma_\zeta d(A,B)^\zeta}.
\end{eqnarray}
where we've redefined our constants. This is the polylogarithmic light cone
for the two-point dynamical correlation functions. Compare this to
the original Lieb-Robinson bound, which reads
\begin{equation}
  \left\| \left[ A , B(t) \right] \right\| \leq
  c e^{k_1|t|} e^{-k_2d(A,B)}.
\end{equation}

To conclude we'd like to mention two consequences of our
light cone for the disordered $XY$ model. The first is a proof of
the conjecture that two parties, Alice and Bob, with access to only
bounded regions $A$ and $B$ at either end of the chain,
respectively, cannot use the dynamics of the disordered model to
send information from Alice to Bob. We follow the argument of
\cite{bravyi:2006a}, appropriately modified to take account of our
stronger bound.

Let $C = L\setminus (A\cup B)$, where $L$ is the chain, be the
region that Alice and Bob cannot access. The most general way Alice
can encode her message is via a set of unitary operators
$\{U_A^k\,|\, k=1, 2, \ldots, m\}$ on her system, where $k$ is
varied according to the message she wants to send. After a time $t$
has elapsed the system has evolved from an initial state $\rho_0$ to
$\rho(t) = e^{-iHt}\rho_0e^{iHt}$. We interpret this as a quantum
channel with input $\rho_{ABC}^k = U_A^k\rho_0 {U_{A}^k}^\dag$ and
output $\rho_{B}^k(t) = \tr_{AC}(U_A^k(t)\rho_0{U_{A}^k}^\dag(t))$.
As argued in \cite{bravyi:2006a}, the output states are all very
close together, as measured in trace norm:
\begin{equation*}
\|\rho_{B}^k(t) - \rho_{B}(t)\|_1 \le c_\zeta n^2 |t|
e^{-\gamma_\zeta d(A,B)^\zeta},
\end{equation*}
where $\rho_{B}(t) = \tr_{AC}(e^{-iHt}\rho_0e^{iHt})$.

If Alice applies the unitaries $\{U_A^k\}$ according to the
probability distribution $\{p_k\}$, the amount of information that
is sent through the channel is given by the Holevo capacity:
\begin{equation*}
\chi(t) = S\left(\sum_{k=1}^m p_k \rho_B^k(t)\right) - \sum_{k=1}^m
p_k S(\rho_B^k(t)),
\end{equation*}
where $S(\cdot)$ is the von Neumann entropy. Applying Fannes
inequality \cite{nielsen:2000a} we find that
\begin{equation*}
\chi(t) \le 2\epsilon(|B|-\log_2(\epsilon)),
\end{equation*}
where $\epsilon = c_\zeta n^2 |t|e^{-\gamma_\zeta d(A,B)^\zeta}$. That is, Bob has
to wait a subexponentially long time (in $d(A,B)$) before a nontrivial
amount of information can arrive. The optimal encoding for Alice to
adopt was investigated in \cite{osborne:2003b} and
\cite{burgarth:2005a} and completely solved in the single-use case
in \cite{haselgrove:2005a}.

The second consequence of the polylogarithmic light cone bound is that
the entropy of any contiguous block $B$ of spins in a dynamically
evolving product state $|\psi(t)\rangle = e^{itH}|00\cdots 0\rangle$
is bounded. Indeed, applying the argument of \cite{eisert:2006b,
grimmett:2007a}, we find that $S(\rho_B(t)) \le c_1 +
c_2\log_2^{\frac{1}{\zeta}}(n|t|)$ as $|B| \rightarrow \infty$, where $c_1$ and $c_2$
are constants. This provides a theoretical explanation for the
phenomenon numerically observed by De Chiara \emph{et.\ al.}
\cite{dechiara:2006a}. It seems nontrivial to adapt the argument of
\cite{bravyi:2006a} to prove the same result because their proof
can't be simply modified to make use of the presence of disorder.

{\it Acknowledgments---}Helpful conversations with Jens Eisert are
gratefully acknowledged. This work was supported by the EPSRC and
the Nuffield Foundation.

\end{document}